\documentclass[
    ,final            % use final for the camera ready runs
  ]
  {aipproc}

\layoutstyle{8x11single}

\begin{document}

\title{Finding the Charge of the top quark in the Dilepton Channel}


\author{A. Beretvas}{
  address={Fermilab/CDF}
}

\author{J. Antos}{
  address={Academia Sinica/Institue of Experimental Physics,Kosice Slovakia}
}

\author{Y.C. Chen}{
  address={Academia Sinica}
}

\author{ Z. Gunay, V. Sorin, K. Tollefson}{
  address={Michigan State University}
}
  
\author{P. Bednar, S. Tokar}{
  address={JINR Dubna/Comenius U. Bratislava}
}

\author{V. Boisvert, W. Hopkins, K. McFarland}{
  address={University of Rochester}
}

\begin{abstract}
There is a question about the identity of the top quark. Is it the top
 quark of the Standard Model (SM) with electric charge $\frac{2}{3}$ or is it an
 exotic quark with charge -$\frac{4}{3}$? An exotic quark has been proposed
 by D. Chang $et$ $al.$\cite{Exotic}. This analysis will use the standard
 CDF run II dilepton sample. The key ingredients of this analysis are the
 correct pairing of the lepton and b-jet, the determination of the charge of
 the b-jet. The analysis proceeds by using
 a binomial distribution and is formulated so that rejecting one hypothesis
 means support for the other hypothesis.
\end{abstract}

\maketitle

%%%%%%%%%%%%%%%%%%%%%%%%%%%%%%%%%%%%%%%%%%%%
%% MAINMATTER
%%%%%%%%%%%%%%%%%%%%%%%%%%%%%%%%%%%%%%%%%%%%

\section{Introduction}
In the SM the top quark decays immediately into $t$ $\rightarrow$ $W^{+}b$,
while in the Exotic Model (EM) into $t$ $\rightarrow$ $W^{-}b$ (and cc).
We measure the charge of the lepton which is the same as that of it's parent ($W$). 
In the top model of D. Chang
$et$ $al.$ the exotic quark is part of a doublet with the other quark ("top")
having a mass of 230 GeV or higher. We are interested in establishing whether the
quark we observe(top) has electric charge $\frac{2}{3}$ or -$\frac{4}{3}$ independent of
any model. The signature for the SM is that the lepton and the associated $b$ have opposite 
charges (either -+ or +-), while in the (EM) they have the
 same charges (either ++ or --).

We plan to look at the dilepton sample for an integrated luminosity = 1
fb$^{-1}$. The present dilepton sample contains 64 events and corresponds to an
integrated luminosity of 0.75 fb$^{-1}$.

\section{Key Ingredients}
There are two key ingredients to the method. First, we need to know how to do the
pairing between the lepton
and the b-jet. The top mass has been determined using a kinematic
method \cite{Lysak}. In the kinematic method a set of equations is obtained
from the measured momentum of the b-quarks and leptons, the two components of
the measured missing E$_{t}$, and assumptions about the six final-state
particles masses, an additional constraint on the longitudinal momentum of
the $t\overline{t}$ system, and constraints on the decays of the $W$, $t$, 
and $\overline{t}$. The system in general has 8 solutions, at the final stage
there are two solutions left, we choose the solution which has the largest number
of entries as corresponding to the correct pairing. Second, we need to
 find the charge of the $b$-jet.  
We determine the charge of the b-quark by looking at the charge of the particles
coming from the $b$-jet. The following equation is used:
\begin{equation}
 Q_{\rm jet} = \frac{\sum_{tracks}q_{i}(\overrightarrow{P_{i}} \cdot \overrightarrow{P_{\rm jet}}  )^{0.5} }
 {\sum_{tracks}(\overrightarrow{P_{i}} \cdot \overrightarrow{ P_{\rm jet}}  )^{0.5} }
\end{equation}

\section{Charge Symmetry}
We will refer to the reconstructed charge distribution as P$_{b}$(x).
The corresponding symmetry relation is P$_{b}$ = P$_{\overline{b}}$(-x).
We use the variable $t_{Q}$ = $\mid$ x+l$_{Q}$ $\mid$ where x is the reconstructed b-jet
charge and $l_{Q}$ is the charge of the paired lepton. 
In Fig. 1 (the x-axis is Q$_{jet}$ as defined by Eq. 1)
we show the $b$ ($\overline{b}$) charge distribution obtained from the Pythia Monte Carlo
for a top mass of 175 GeV/$c^{2}$.

%%%%%%%%%%%%%%%%%%%%%%%%%%%%%%%%%%%%%%%%%%%%
%% Sample figure:
%%
%% The option [height=...] scales the picture to the given height,
%% without it it would be printed at its nominal size
%%%%%%%%%%%%%%%%%%%%%%%%%%%%%%%%%%%%%%%%%%%%

\begin{figure}
  \includegraphics[height=.3\textheight]{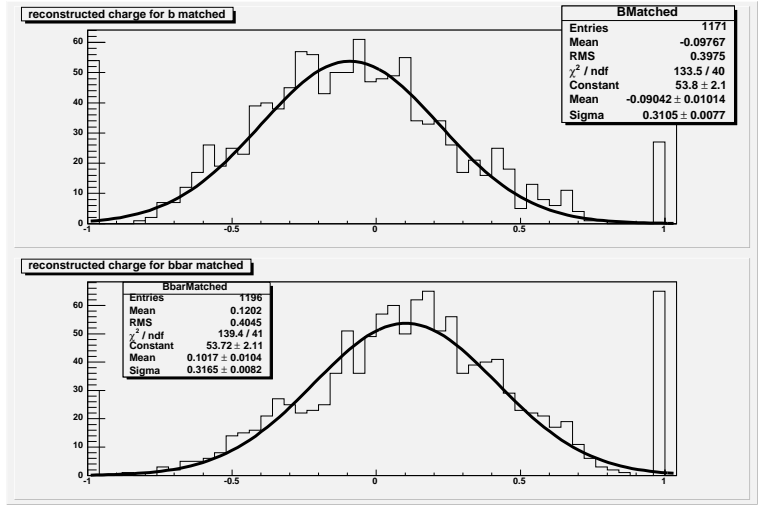}
  \caption{Reconstructed $b (\overline{b})$ jet charge distribution.}
\end{figure}

\section{Basic Equation}
The charge of the top quark is related to the correct pairing of the lepton and
the b-quark and knowing the charge probability distribution of the b-jet.
P$_{lb}$ is the probability for the correct pairing of the lepton and the b-jet.
When the pairing is always correct the probability distribution of the variable
$t_{Q}$ is related to the probability distribution of the b-quark: P($t_{Q}$) =
P$_{b}(t_{Q} -1)$. This can be generalized to:
\begin{equation}
 {\rm P}(t_{Q}) = {\rm P}_{lb}{\rm P}_{b}(t_{Q}-1) +
 (1 - {\rm P}_{lb}){\rm P}_{b}(1 - t_{Q}).
\end{equation}
The corresponding equation for the exotic hypothesis is:
\begin{equation}
{\rm P}^{Ex}(t_{Q}) = {\rm P}_{lb}{\rm P}_{b}(1 - t_{Q}) +
 (1 - {\rm P}_{lb}){\rm P}_{b}(t_{Q}-1).
\end{equation}
We formulate the problem of determining the charge of the top quark as a 
problem in binomial statistics. We define the charge of the b-jet to be
-$\frac{1}{3}$ if the reconstructed b-jet ($\overline{b}$-jet)
charge is less(greater) than 0. For the 
Pythia charge distribution shown in Fig.1 this is the correct determination 63.0
$\pm$ 1.3\% of the time. The problem of correctly associating the $b$-jet and
and the lepton is a problem that we have solved in our kinematic analysis of
the $t$ quark mass. The smeared distribution of the two possible pairings are
compared and the one with the largest number of entries is selected. The correct
solution is found 70\% of the time.

	We define P$^{++}$ to be the probability to measure top charge
	+$\frac{2}{3}$ given that the SM is true and P$^{+-}$ to be the
	probability to measure top charge -$\frac{4}{3}$ given the SM is true.
	For a perfect experiment P$^{+-}$ would be 0.
\begin{equation}
    {\rm P}^{++} = {\rm P}^{charge}_{b}{\rm P}_{lb} + (1-{\rm P}^{charge}_{b})(1-{\rm P}_{lb})	
\end{equation}

For the numbers given we have P$^{++}$ = 0.63$\times$0.7 +(1-0.63)(1-0.7) = 0.55.
%%%%%%%%%%%%%%%%%%%%%%%%%%%%%%%%%%%%%%%%%%%%
%% SAMPLE TABLE
%%
%% Shows the use of \tablehead and \tablenote
%% macros
%%%%%%%%%%%%%%%%%%%%%%%%%%%%%%%%%%%%%%%%%%%%
\section{Binomial Statistics}
The probability to observe N$^{++}$ or less pairs out of N pairs (where each
pair contains a lepton and $b$-jet which have opposite charge) is:
\begin{equation}
{\rm P}({\rm N}^{++}) = \sum_{i=0}^{{\rm N}^{++}}\left( \begin{array}{c} {\rm N} \\ i \end{array} \right)
({\rm P}^{++})^{i} (1 -{\rm P}^{++})^{{\rm N}-i}
\end{equation}

\section{Backgrounds}
We have verified that the probability for measuring $t_{Q}$ in background
events is 50\% by running on Monte Carlo events. The three background processes are Drell-Yan, $W$ $\rightarrow$
$l\nu$ + jets (where the jet fakes a lepton) and Dibosons. Our formula is
modified to include backgrounds:
\begin{equation}
{\rm P}^{++}_{bkg} = {\rm P}^{++}f_{top} + 0.5f_{bkg}
\end{equation}

Our expectation is that 62\% of the events will be signal and 38\% background.
Thus P$^{++}_{bkg}$ = 0.55$\times$0.62 + 0.5$\times$0.38 = 0.53

\section{Improving the lepton $b$ Pairing}
The square of the invariant mass of the lepton and the b can be easily
calculated. There are 4 possible values as there are two leptons and two jets.
Two of the combinations are associated with the correct pairing and two with
the incorrect pairing. By running a Monte Carlo we can see which pairings
are correct and which are incorrect. It turns out that only incorrect
pairing are observed at very high values of the invariant mass. This means the other
solution is the correct one.
Using a cut value of 22000 GeV$^{2}$/$c^{4}$ allows one to select
a sample that has a 97\% probability of being the correct pairing.
The efficiency of this cut is only 40\%. We call this method M$^{2}_{lb-max}$.

\section{Figure of Merit}
How can we compare different methods? For example is it better to use the
kinematic method or the M$^{2}_{lb-max}$? The figure of merit is 
$\epsilon$D$^{2}$. The efficiency is $\epsilon$ and the Dilution is D.
The Dilution D = 2$\times$Purity -1. Thus the figure of merit for the
M$^{2}_{lb-max}$ method is 0.40$\times$(2$\times$0.97 -1)$^{2}$ = 0.35, and the figure
of merit for the kinematic method is 1.0$\times$(2$\times$0.7 -1)$^{2}$ = 0.16.
For the kinematic method the efficiency is much higher, but the purity is lower
and we find the M$^{2}_{lb-max}$ is more than a factor of 2 better.

 \section{Plans}
The most difficult part of the analysis will be to use Calibration data
(QCD  $b\overline{b}$ dijets) to obtain b-jet charge distribution and to
extrapolate these data so that we understand the $b$-jet charge distribution
at high p$_{t}$. It will be interesting to compare these results with Fig.1.
We plan to make measurements
in the dilepton channel  and also in the lepton plus jet channel.

%%%%%%%%%%%%%%%%%%%%%%%%%%%%%%%%%%%%%%%%%%%%%%%%
%% BACKMATTER
%%%%%%%%%%%%%%%%%%%%%%%%%%%%%%%%%%%%%%%%%%%%%%%%

\begin{theacknowledgments}
We wish to thank all members of the CDF top group for their help in making this
analysis possible.
  \end{theacknowledgments}



\bibliographystyle{aipproc}   % if natbib is available

\bibliography{sample}

\IfFileExists{\jobname.bbl}{}
 {\typeout{}
  \typeout{******************************************}
  \typeout{** Please run "bibtex \jobname" to optain}
  \typeout{** the bibliography and then re-run LaTeX}
  \typeout{** twice to fix the references!}
  \typeout{******************************************}
  \typeout{}
 }

%%%%%%%%%%%%%%%%%%%%%%%%%%%%%%%%%%%%%%%%%%%
%% The following lines show an example how to produce a bibliography
%% without the help of the BibTeX program. This could be used instead
%% of the above.
%%%%%%%%%%%%%%%%%%%%%%%%%%%%%%%%%%%%%%%%%%%

\end{document}